\documentclass[prc,preprintnumbers,twocolumn,showpacs,nofootinbib,aps,longbibliography,superscriptaddress]{revtex4-1}
\usepackage[utf8]{inputenc}
\usepackage[T1]{fontenc}
\usepackage{color}
\usepackage{amsmath}
\usepackage{url}
\usepackage{graphicx}
\usepackage{amssymb}
\usepackage{appendix}
\usepackage[section]{placeins}
\usepackage{multirow}
\usepackage{hyperref}
\usepackage{siunitx}
\usepackage{setspace}
\usepackage{mathrsfs} 
\usepackage{bm}
\usepackage{xcolor}
\usepackage{soul}
\usepackage{array}
\usepackage{lipsum}
\usepackage{relsize}
\usepackage{amsfonts}
\usepackage{comment}
\usepackage{outlines}
\usepackage[normalem]{ulem}
\usepackage{multirow}
\usepackage{caption} 
\usepackage{subcaption} 
\usepackage{braket}
\usepackage[printonlyused,nolist]{acronym}
\usepackage[bb=boondox]{mathalfa}
\usepackage{footnote}
\usepackage{academicons}
\usepackage{xcolor}
\usepackage[TU]{fontenc}
\usepackage{MnSymbol}
\definecolor{orcidlogocol}{HTML}{A6CE39}

\begin{document}
\title{Uncertainty quantification of optical models in fission fragment de-excitation}


\author{Kyle A. Beyer \footnote{a}}
 \email{beyerk@frib.msu.edu}
 \homepage{work completed as part of dissertation work at $^2$}
\affiliation{%
Facility for Rare Isotope Beams, Michigan State University, East Lansing, Michigan 48824, USA
}%
\affiliation{
Department of Nuclear Engineering and Radiological Sciences, University of Michigan, Ann Arbor, MI 48109, USA
}%

\author{Amy E. Lovell}%
\affiliation{%
Theoretical Division, Los Alamos National Laboratory, Los Alamos, NM 87545, USA
}%

\author{Cole D. Pruitt}
\affiliation{
Lawrence Livermore National Laboratory, Livermore, CA 94550, USA
}%

\author{Nathan P. Giha}
\author{Brian C. Kiedrowski}
\affiliation{
Department of Nuclear Engineering and Radiological Sciences, University of Michigan, Ann Arbor, MI 48109, USA
}%

\date{\today}

\preprint{LA-UR-23-31907}

\begin{abstract}
We take the first step towards incorporating compound nuclear observables at astrophysically-relevant energies into the experimental evidence used to constrain optical models, by propagating the uncertainty in two global optical potentials, one phenomenological and one microscopic, to correlated fission observables using the Monte Carlo Hauser-Feshbach formalism. We compare to a wide range of historic and recent experimental fission measurements, and discuss in detail regions of disagreement. We find that the parametric optical model uncertainty in neutron-fragment correlated observables involving neutron energy is significant. On the other hand, we observe that other experimental features, particularly neutron-fragment correlations near the $^{132}$Sn shell-closure and the high energy component of neutron spectra, are unlikely to be explained by the optical potential, and will require further experimental and theoretical effort to explain.  
\end{abstract}

\maketitle

\section{Introduction}

How nucleons organize themselves across the nuclear chart, including the origin of heavy isotopes in the astrophysical r-process, requires understanding of nuclear reaction phenomenology away from stability \cite{hebborn2022,osti_1296778}. The extrapolation of reaction models into the region of fission fragments is specifically necessary for predictive cross sections required for r-process simulations, nuclear non-proliferation and forensics applications and for modeling the fission process itself. In particular, accurate fission event generators are important to nuclear science and technology, for the interpretation of experimental fission observables, and for evaluation of \ac{PFNS} for actinides \cite{trkov2015current, capote2016prompt}. \ac{MCHF} codes are commonly employed in these efforts, as fission produces multiple correlated signatures \cite{lovell2021correcting, lovell2022energy,talou2021fission}.  

\begin{figure*}
\begin{center}
  \includegraphics[width=\textwidth]{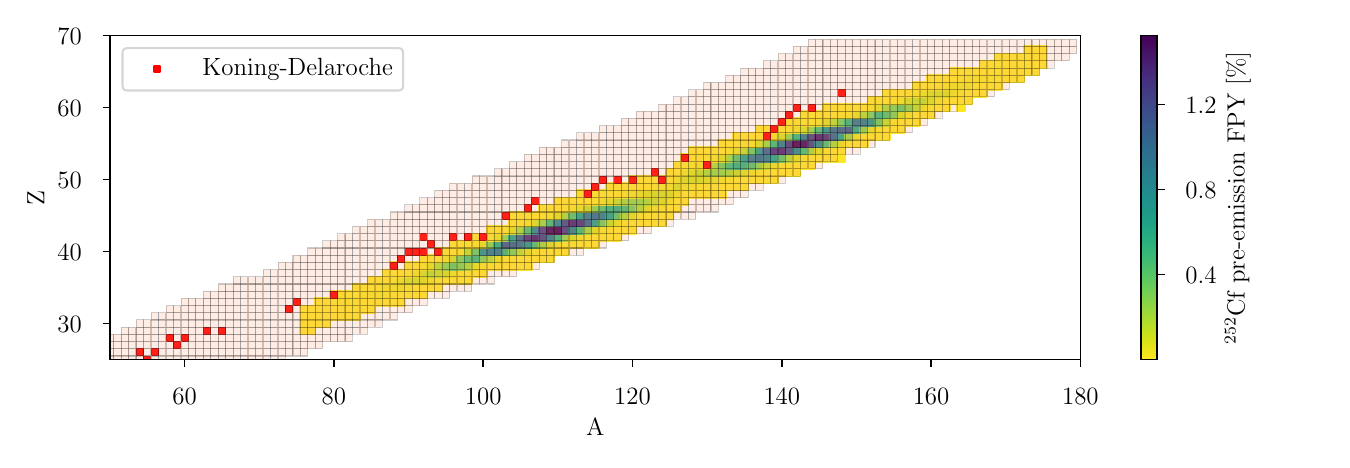}
\end{center}
\caption{Chart of isotopes in the fission fragment region, where the isotopes used to constrain the commonly used Koning-Delaroche phenomenological neutron-nucleus global optical potential \cite{koning2003local} (red squares) are compared to $^{252}$Cf$(sf)$ pre-emission fragment yield distributions, as generated by \acs{CGMF} \cite{talou2021fission}.
\label{fig:coi}
}
\end{figure*}

The typical workflow for nuclear reaction phenomenology often involves constraining a global nucleon-nucleus \ac{OMP} to elastic scattering experiments on $\beta$-stable targets, and then extrapolating away from stability to the neutron-rich region, for which experimental data are unreliable or non-existent. Global phenomenological \ac{OMP}s provide workhorse models for the nucleon-nucleus interactions that are key ingredients in the formalism for direct and compound reactions. However, extrapolating away from the isotopic and energy regions for which they are fit introduces unquantified uncertainty. This need for extrapolation is exemplified in Fig.~\ref{fig:coi}, in which the isotopes used to constrain a workhorse global phenomenological optical potential \cite{koning2003local} are compared to fission fragment yields.

In particular, the imaginary surface component of the optical potential is peaked at low energies, reflecting that low-energy neutrons are preferentially absorbed, and emitted, from the fringe regions of the nucleus, where nucleon density, and, therefore, Pauli blocking effects, are diminished relative to the interior \cite{moldauer1962optical}. Understanding the isovector ($\alpha = \pm\frac{N-Z}{N+Z}$) dependence of this low-energy imaginary surface strength, where the plus (minus) sign is for incident protons (neutrons), is essential for astrophysical applications, with simulations predicting a strong sensitivity of r-process reaction rates to the isovector dependence, especially for drip-line nuclei \cite{goriely2007isovector}. 

A key ingredient in nuclear reaction modeling away from stability are global \ac{OMP}s, the Hauser-Feshbach formalism for decay chains of compound nuclei being no exception. This motivates the quest for predictive phenomenology away from the line of stability. The advent of the rare isotope beam era has begun to widen the experimentally accessible region of the isotopic chart, with significant increases still to come \cite{bollen2010frib}. The development of uncertainty-quantified \ac{OMP}s that are predictive for these rare isotopes has been the subject of a recent corresponding theoretical effort \cite{hebborn2022}. Simultaneously, there has been a focus on applying rigorous uncertainty quantification using Bayesian statistics to the calibration of phenomenological optical potentials \cite{PhysRevC.97.064612,PhysRevLett.122.232502, PhysRevC.100.064615,lovell2020recent, catacora2021statistical, pruitt2023uncertainty}.

Most of these efforts calibrate the optical potential parameters to elastic scattering experiments. The corpora of experimental data typically include include neutron and proton differential elastic scattering cross sections, $d\sigma/d\Omega$, and total and reaction cross sections, $\sigma_t$ and $\sigma_r$, respectively, as well as analyzing powers. Current work has focused on propagating \ac{OMP} uncertainties into specific non-elastic reaction channels, such as charge-exchange and knockout reactions, with the goal of eventually including non-elastic measurements as \ac{OMP} constraints in the form of Bayesian posteriors for model calibration \cite{whitehead2022prediction,hebborn2022quantifying,smith2024uncertainty}. 

In this work, we propose the expansion of this corpus to include \ac{CN} observables in which unstable nuclei are produced, directly into the calibration likelihood for \ac{OMP}s. Specifically, we focus on observables from fission. This has several potential advantages:

\begin{enumerate}
\item measurements of fission fragments and their prompt emissions provide experimental access to unstable nuclei, thereby directly testing the extrapolations of global \ac{OMP}s from stability, 
\item the energy scale of prompt neutron emissions is typically closer to astrophysically relevant energies than direct reactions in beam-line experiments,
\item and there already exists a large library of historical experimental data for fission-fragment decay spectroscopy, including correlations
\end{enumerate}

It is also worth noting that \ac{CN} processes already contaminate elastic scattering measurements at low incident nucleon energy, where compound-elastic processes dominate the cross sections.  In one way or another, all phenomenological \ac{OMP}s that extend to low energies must address this in the calibration. Due to computational limitations, the approximation may be adopted to pre-calculate the \ac{CN}-elastic cross section with a \ac{HF} treatment, using a single set of \ac{OMP} parameters from the prior; that is, the \ac{CN}-elastic component is typically not adjusted with the calibration \cite{pruitt2023uncertainty}. 

Directly incorporating \ac{CN} observables, including fission, potentially provides new constraints for phenomenological \ac{OMP}s at low-energy, that may be under constrained currently. The sensitivity of \ac{CN} reaction formalisms to model inputs has been tested in the past; e.g. for  non-statistical properties \cite{PhysRevC.104.014611} in fission, and for level densities in fusion \cite{voinov2007test}. 

It is a computationally demanding task to simultaneously quantify uncertainty and correlations between all model inputs for every \ac{CN} observable of interest, but the piecemeal efforts so far have indicated strong sensitivity to level densities and cases where results disagree significantly with experiment \cite{voinov2007test}. This indicates the potential efficacy of optimizing model inputs --- level densities, \ac{OMP}s, strength functions, etc. --- to \ac{CN} observables; and, indeed, level densities have previously been extracted from neutron evaporation spectra, \cite{wallner1995extraction,ramirez2013nuclear} to name a few. It has been suggested that uncertainties in \ac{CN} observables due to the \ac{OMP} may be larger than for level densities in some cases; particularly at low energy \cite{pruitt2023uncertainty}. Therefore, we take the first steps here to quantify this uncertainty in fission.

Nuclear fission has been a subject of study since its discovery by Lise Meitner and Otto Frisch in the experimental results of Otto Hahn in 1938. The phenomenon, in which a fissile nuclide - typically an actinide - deforms into an unstable configuration which splits into (typically) two fragments, produces a variety of correlated observables, and spans many orders of magnitude in time. 

Due to the interesting, still not fully understood, physical phenomena associated with fission, and the importance of the reaction to society, it has been subject to extensive measurement and modeling (For an exhaustive review see \cite{talou2023nuclear, schunck2022theory} and references therein). The observables we consider in particular are the prompt neutrons emitted from the fragments as the de-excite, which are strongly correlated with the excitation energy and angular momentum of the fragments following scission, as well as the branching ratios for de-excitation. 

There is a rich history of experiments investigating the properties and correlations between the prompt emissions in fission going back at least six decades \cite{schmitt1965precision,marin2020event,oberstedt2013improved,wilson2021angular, hoffman1964directional, franklyn1978angular, BUDTZJORGENSEN1988307}. Measured observables include the energy and multiplicity of neutrons and photons, and the correlations between them and the fragments themselves.  Fission experiments present a breadth of data which can potentially constrain \ac{OMP}s.

Simultaneously, it is desirable to understand the model dependence of predictions of fission codes. The pre-emission fragment states cannot be directly measured, and the effect of the dependence on the de-excitation model of reconstructions of the pre-fragment state must be well understood to use experiments to probe open questions about excitation energy and angular momentum sharing in fission \cite{wilson2021angular}. Specifically, it has been observed that \ac{OMP} predictions of the removed
angular momenta of emitted neutrons is larger than expected \cite{stetcu2021angular}. 
{
Understanding the role of the \ac{OMP} with rigorous uncertainty quantification allows for, not only the identification of observables that are sensitive to the \ac{OMP} and may be used to constrain it, but also for the identification of observables that are not, and may be useful to constrain other aspects of fission, e.g. the energy and angular momentum of the fragments immediately following
scission.}

The post-scission fragment is born into a state dependent on the conserved quantities (mass, charge, angular momentum, excitation energy, etc.) of the pre-scission compound nucleus, its deformation at the saddle point, and the dynamics of scission. It then equilibrates its excess deformation energy into excitation energy and, typically, promptly emits a neutron or two.  Once its residual excitation  energy approaches the neutron separation energy, photon emission will strongly compete with
neutrons, as the fragment continues to de-excite through statistically distributed levels. Once the excitation energy and parity of the fragment reaches the Yrast line, discrete photons are emitted, carrying away the majority of the fragments excess angular momentum. Following this prompt de-excitation, the residual fragment will often remain $\beta$ unstable, leading to further delayed emissions and long-lived products. 

It is the goal of this work to determine to what degree measurements of the prompt emissions provide constraints on the nucleon-nucleus interaction between the residual fragments and their neutron progeny.  We investigate this dependence for the first time by propagating two uncertainty-quantified \ac{OMP}s through \ac{MCHF} fission fragment de-excitation with the \acs{CGMF} code, and constructing credible distributions for a variety of fission observables. These distributions represent the uncertainty of a given fission observable, due to the parametric uncertainty in an \ac{OMP}. 

The paper is organized as follows: in section ~\ref{sec:theory}, we briefly review the theory of nucleon evaporation from an excited nucleus. In section ~\ref{sec:uq}, we discuss the implementation of this theory in \acs{CGMF} and the observables considered, and in ~\ref{sec:results}, we report and discuss results, discussing in detail any disagreements between the different \ac{OMP}s and theory. Finally, in section ~\ref{sec:conclusion}, we comment on the sensitivity of fission observables to \ac{OMP}s, as well as discuss future work directly incorporating these observables as constraints, and in the global optimization of model inputs into \acs{CGMF}, enumerating the assumptions required to do so.

\section{Theory of nucleon emission from an excited nucleus}
\label{sec:theory}

Calibrating an optical potential to \ac{CN} observables of a given reaction is reasonable to propose if (1) the \ac{HF} formalism (or an extension of it) sufficiently describes the reaction and (2) the reaction explores a phase space (e.g. in mass, charge and energy) not available in typical elastic scattering experiments. As discussed, the latter is clearly true for fission. In this section we discuss the former; briefly reviewing the \ac{HF} theory of the \ac{CN} to clarify the connection between the \ac{OMP} and the decay of the \ac{CN}, and clearly enumerate the assumptions that are made so that this connection may be utilized only when applicable. The bulk of the formalism developed in this section is included in standard textbooks on reaction theory, e.g. \cite{thompson2009nuclear}, but we specialize the discussion to its application to fission fragments.

Prompt neutron emission is governed by the initial state of the fragment after scission, as well as the level density of the residual core, and the matrix element for the emission. In particular, consider the state of the emitted nucleon, with asymptotic kinetic energy $E = \hbar^2 k^2 / 2 \mu$:

\begin{equation}
  \begin{split}
  \ket{kjl} &\equiv \ket{k} \sum_m \mathcal{Y}^m_{jl}  \\
            &= \ket{k} \sum_{m_s m_l} i^l \braket{l s m_l m_s | j m}  Y_l^{m_l} \chi_{s}^{m_s}
  \end{split}
\end{equation}

\noindent
with $\mu$ being the reduced mass of the system. Here, $l$, $s = 1/2$ and $j$ respectively label the orbital, spin and total angular momentum quantum numbers, with $m_l$, $m_s$ and $m$ their respective projections. $\chi_{s}^{m_s}$ is the spin wavefunction, $Y_l^m$ are the orbital wavefunctions, e.g., the spherical harmonics, and $\mathcal{Y}_{l}^{m_l}$ are the corresponding spin-orbital angular wavefunctions.  

The probability of emitting a neutron described by this state from a \ac{CN} system with total spin $I$ and parity $\pi$ is given by the transmission coefficient, $\mathcal{T}_{jl}^{I \pi}$, which is defined as 

\begin{equation}
  T_{jl}^{I \pi}(E) = 1 - |\mathcal{S}_{jl}^{I \pi}(E)|^2,
\end{equation}

\noindent 
where the partial wave $\mathcal{S}$-matrix is due to the effective interaction between the neutron and the residual core, the \ac{OMP}.

Our scenario of interest is an initial $A+1$ nucleus in some eigenstate $m$ with energy $E_m^{A+1}$ and total angular momentum $I_m$ emitting a nucleon to leave the residual $A$-body core in a state $n$ with energy $E_n^A$ and total angular momentum $I_n$. The neutron-removal threshold is $S^{A+1}_N = E_0^{A+1} - E_0^A$, where $n=0$ refers to the ground state of the respective system. Of course, energy, parity and total angular momentum are symmetries of the \ac{COM}-frame Hamiltonian, so we have a triangular sum rule for total angular momenta:

\begin{equation}
  \label{eq:triangle}
  |I_n - j| < I_m < I_n + j,
\end{equation}

\noindent
as well as conservation of energy: neutron emission with energy $E$ in \ac{COM}-frame, leaves the excited residual core with energy $E_n^{A}$, such that 

\begin{equation}
  \begin{split}
    E_{m}^{A+1} &= E_n^{A} + E  \\
    E_{x}^{A+1} - E_{x'}^A + S_N^{A+1} &= E,
  \end{split}
\end{equation}

\noindent
where $E_x$ refers to excitation energy above the corresponding ground state.

The formula for the probability (or fractional width) for the compound nucleus to de-excite through this channel is the ratio of the corresponding transmission coefficient to the sum of those to all open channels: 

\begin{equation}
  \begin{split}
    \label{eq:HF}
    p(\alpha) &= \frac{\Gamma_{\alpha}}{\sum_\beta \Gamma_\beta} \equiv \frac{\Gamma_\alpha}{\Gamma}\\
             &= \frac{\mathcal{T}_{\alpha} }{ \sum_{\beta} \mathcal{T}_{\beta} } .
  \end{split}
\end{equation}

This is the Hauser-Feshbach theory \cite{hauser1952inelastic}. Here $\Gamma_\alpha$ refers to the $\alpha$ channel decay width, which is proportional to the corresponding transmission coefficient (up to normalization), neglecting \ac{WFC} factors (e.g. assuming no correlation between the entrance and exit channel) \cite{SATCHLER196355}. 

{
  A discussion on the validity of this assumption is warranted. If neutrons are removed immediately following scission before the thermalization or full acceleration of the fragments, then they could in principle be correlated with the process of scission itself - in other words neutrons removed from the fragments would have different energy or angular distributions than neutrons removed from the same systems at the same energies if you were to prepare them some other way besides through scission. If this were the case, one could in principle generalize the description of \ac{WFC} factors to include correlations between the dynamics of scission and this post-scission, pre-equilibrium neutron emission from the fragments. A similar argument can be made for subsequent neutrons emitted from a parent fragment and its daughter(s).
}

{
  Simply comparing the expected time scales of scission ($10^{-22}$s) and prompt emission ($10^{-18}$s) indicates that the fission fragments are fully thermalized between scission and prompt emission, and between subsequent prompt emissions, removing any correlations. Therefore, it is correct to ignore \ac{WFC} in \ac{MCHF} de-excitation of fission fragments.
}

Where we do not have information on individual discrete levels, we turn the sum in Eq.~\eqref{eq:HF}, which is over channels $n$ with definite \ac{COM}-frame energy $E_n$, into an integral 

\begin{equation}
  \begin{split}
    \sum_n \mathcal{T}(E=E_m^{A+1}-E_n^A) 
    &\rightarrow 
    \int dE_{n}^A \left( \frac{dn}{dE_n^A} \right)
    \mathcal{T}(E_m^{A+1} - E_n^A) \\
    &= \int dE_{x}^A \underbrace{\left( \frac{dn}{dE_x^A} \right)}_{\rho(E_{x}^A)} 
    \mathcal{T}(E) ,
  \end{split}
\end{equation}

\noindent
which defines the level density $\rho(E_x^A) dE_x^A$ of the residual core as a function of excitation energy above the core ground state. We have, by conservation of energy, the residual excitation energy of the core $E_x^A = E_x^{A+1} + S_N^{A+1} - E$. This allows us to re-write Eq.~\eqref{eq:HF}, expressing the fractional width for neutron evaporation as
\vspace{12mm} 
\begin{widetext}
\begin{equation}
  \label{eq:HFfull}
  p(E,j,I_n,\pi_n) = \frac{ \mathcal{T}_{jl}^{I_n \pi_n}(E) \; \rho_{jl}^{I_n \pi_n}( E_x^{A+1} + S_N^{A+1} - E)}
  { \sum_{j'I'\pi'} \int dE' \; \mathcal{T}_{j'l'}^{I'\pi'}(E') \;
  \rho_{j'l'}^{I'\pi'}(E_x^{A+1} + S_N^{A+1} - E') }.
\end{equation}
  
\end{widetext}
  
 This is the central result of the Hauser-Feshbach theory, and is used to model the emission of neutrons from an excited \ac{CN}. Here, the superscript $I\pi$ implies the limitation to states $jl$ in the subscript that satisfy the triangle relation Eq.~\eqref{eq:triangle}, and conserve parity.  Resolved discrete excited states of the residual core may be included by keeping them in a discrete sum. In practice, the continuum is discretized into uncoupled bins. The sum in the denominator runs over all open channels that respect the symmetries of the system. Equation ~\eqref{eq:HFfull} can also include photon emission in full competition with neutrons, with $\mathcal{T}_{jl}^{I\pi}$ also representing the photon transmission coefficient, e.g. as calculated according to giant resonance parameters (see \cite{talou2021fission}). 

One can use Eq.~\eqref{eq:HFfull} to model the decay chain of compound nuclei. Decay by nucleon emission requires as ingredients only the level density of the residual core, and the \ac{OMP} that describes the effective interaction between the residual core and the emitted nucleon. 

In this way, given an initial nucleus with well-determined excitation energy, spin and parity, one can compute the probability to decay via a specific mode to a residual nucleus with another excitation energy, spin and parity, and an emitted species of radiation. In fact, the probability in Eq.~\eqref{eq:HFfull} defines a Markov process, in which the current state of the system (e.g. $A,Z,E,J,\pi$) solely determines its probability of evolving to the next state. By sampling from this probability, one can generate an ensemble of \textit{histories}, each specifying a trajectory of de-excitation, beginning with an initial excited nucleus, and ending in a stable (or long-lived) state. 
From many such histories, one can reconstruct experimental observables relating to the emitted radiation, including full correlations between energy, angle, multiplicity and species, as well as with the remaining stable or long-lived nuclei that are produced. This is called the \acf{MCHF} formalism, and we discuss it more detail in the next section.

\section{Uncertainty propagation in Monte Carlo Hauser Feshbach} 
\label{sec:uq}

In this section we discuss the methodology for generating posterior predictive distributions of fission observables due to two different optical potentials using the \ac{MCHF} method as implemented in the code \acs{CGMF}. The potentials considered in this work are the phenomenological \ac{KDUQ} \cite{pruitt2023uncertainty,koning2003local} and the microscopic \ac{WLH} \cite{whitehead2021global}. The fission reactions under consideration were $^{252}$Cf$(sf)$ and $^{235}$U$(n_{\rm{th}},f)$.

\subsection{Optical potentials}

The \ac{KDUQ} phenomenological global potential \cite{pruitt2023uncertainty} was fit to differential elastic scattering cross sections and analyzing powers, proton reaction cross sections, and neutron total cross sections. It is an uncertainty-quantified version of the workhorse phenomenological \ac{OMP} Koning-Delaroche \cite{koning2003local}, and updates the original potential using a full Bayesian  calibration and outlier rejection. The original potential is valid for 1 \unit{keV} up to 200 \unit{MeV}, for (near-) spherical nuclides in the mass range $24 \leq A \leq 209$. The default parameterization of Koning-Delaroche, as provided in \cite{koning2003local}, is the default \ac{OMP} in \acs{CGMF}. 

In \cite{pruitt2023uncertainty}, two different likelihood models are used, the ``democratic" and the ``federal". In the democratic ansatz, the data covariance in the likelihood function is assumed to be diagonal, the experimentally reported uncertainties for each data point are augmented with an unreported uncertainty (estimated in an observable-by-observable basis) and the covariance is scaled by $k/N$, $k$ being the number of model parameters (46), and $N$ being the total number of data points. In the federal ansatz, the data covariance is modified so that data points of observable $i$ are scaled separately by the number of data points belonging to that observable, e.g., by $k/N_i$. Thus, each observable has equal influence in the likelihood, and each data point has equal influence within its observable. In \cite{pruitt2023uncertainty}, the two ansatze produced similar calibrated models, and the federal ansatz is used in this work.

The \ac{WLH} microscopic global potential was developed in a nuclear matter folding approach \cite{whitehead2021global}. First, in asymmetric nuclear matter, at a variety of densities and asymmetry parameters $\alpha$, the single-nucleon self-energy was calculated self-consistently to 2$^\text{nd}$ order in \ac{MBPT}, using nucleon-nucleon forces from \ac{xeft}. These density-dependent self-energies are calculated for a much wider range of asymmetries than is possible in phenomenological models like \ac{KDUQ}.

The nuclear matter self-energies are then folded to nuclear density distributions using the \ac{I-LDA} to produce optical potentials. The nucleon densities are calculated in mean-field theory with Skyrme effective interactions \cite{PhysRevC.95.065805}. The underlying theoretical uncertainty was approximated by the spread in resulting parameters from five different choices of chiral interaction. The \ac{WLH} spin-orbit term was developed using a density-matrix expansion at the Hartree-Fock level \cite{holt2011nuclear}.  

The targets considered spanned 1800 nuclei, ranging in mass from $12 < A < 242$, inclusive of light and medium-mass bound isotopes out to the neutron drip line. The energy considered was $E \in [0,200]$ \unit{MeV}. 

For target nuclei with small proton-neutron asymmetry, the \ac{WLH} isospin-dependence follows the  Lane form with first-order $\alpha$ dependence, however, for nuclei with larger isospin asymmetries, the \ac{WLH} potential contains terms with no parallel in \ac{KDUQ}, which are proportional to the square of the isospin asymmetry, and strongest at low energy. Understanding the phenomenological implications of these higher order terms is especially interesting for neutron-rich nuclei.

\subsection{\acf{MCHF}}

\acs{CGMF} \cite{talou2021fission} is a \ac{MCHF} code which generates initial configurations for the fission fragments in mass, charge, kinetic energy, excitation energy, spin, and parity, and then samples their de-excitation trajectories by emission of neutrons and $\gamma$ rays according to Eq.~\eqref{eq:HFfull}. It is capable of neutron induced fission on various actinides over incident energies ranging from thermal to 20 \unit{MeV}, as well as spontaneous fission of selected isotopes.

\acs{CGMF} samples the initial post-scission fragment masses using a three-Gaussian model for the distribution in mass number, with energy-dependent centers and standard deviations. The energy dependence is tuned to experimental data. The fragment charges are then sampled conditional on the masses using the Wahl systematics \cite{wahl1980systematics}, and the \ac{TKE} of the fragments are sampled from a Gaussian using a mass-dependent mean and standard deviation tuned to available data otherwise. The \ac{TXE} is then available by energy balance: $\rm{TXE} = Q_f - \rm{TKE}$, where $Q_f$ is the energy liberated in the fission, depending on the binding energies of the pair of fission fragments created. The partitioning of the excitation energy is done in the Fermi-gas model, in which the temperature ratio of the light to heavy fragment is allowed to be mass dependent and tuned to prompt neutron properties. The fragment spins are uncorrelated and chosen from Gaussians centered on the number of geometric levels $J(J+1)$, with width tuned to prompt photon properties. Parity states are assumed to be equiprobable.

Once the post-scission states of the two fission fragments for a history have been sampled, they are then de-excited in the \ac{MCHF} formalism, where the decay of the \ac{CN} is treated as a Markov process, governed by Eq.~\eqref{eq:HFfull}, with neutron and photon emission in full competition. The \ac{MCHF} algorithm, given the conserved quantum numbers of the fully-accelerated post-scission fragment, samples from Eq.~\eqref{eq:HFfull}, emitting neutrons and photons until a state is reached that is stable against prompt emissions. The observables are then re-constructed using the event by event information.

The nuclear level density appearing in Eq.~\eqref{eq:HFfull} is modeled using the \ac{KCK} systematics, which extends the Gilbert-Cameron model to include an energy dependent level density parameter and defines the procedure for assigning spin and parity to discrete levels for which it is unknown \cite{doi:10.1080/18811248.2006.9711062}. Level density models represent another potentially large contributor of uncertainty, as they are plagued by the same issue as \ac{OMP}s of extrapolating away from stability where data is available. 

It is important to note that certain model inputs (e.g. nuclear temperatures used for excitation energy sharing) in \acs{CGMF} have been tuned to experimental observables, especially mean neutron multiplicities. This tuning was performed using the default \ac{KD} \ac{OMP} parameterization, so correlations between these parameters and \ac{OMP} parameters may be responsible for canceling errors or in both. This point will be emphasized in the interpretation of the results.

\subsection{Posterior predictive distributions of fission observables}

For each of the fissioning isotopes considered, the uncertainty propagation was done by brute force, with 300 samples from the posterior distributions of each \ac{OMP}. In this case of \ac{WLH}, these samples were generated by assuming a multi-variate normal distribution and sampling from the covariance provided in \cite{whitehead2021global}. It should be noted that, although the posterior distributions for \ac{WLH} are expected to be well approximated by a multi-variate normal, this approximation may lead to a slightly higher proportion of samples drawn from the tails of the distribution. For the case of \ac{KDUQ}, these samples were taken from the supplemental material of \cite{pruitt2023uncertainty}, using the federal posterior formulation. For each \ac{OMP} sample, an ensemble of one million Monte Carlo histories were generated, from which first and second moments of aggregate observables were re-constructed to compare to experiment. 

The distribution of values for a given fission observable across histories with a single \ac{OMP} sample resulted from the inherent Monte Carlo uncertainty, $\sigma_{mc}$. The distribution of mean values across samples represented an estimate of uncertainty due to the \ac{OMP}, with noise due $\sigma_{mc}$. The uncertainties reported as bands in the results result from the quadrature subtraction of the former from the latter, to estimate the uncertainty due only to the varying \ac{OMP} parameters. For more details, see App.~\ref{ch:appA}.

Within a given ensemble, histories were run in parallel using the MPI implementation in the python module \texttt{mpi4py} \cite{10.5555/898758,DALCIN20051108}. A modified version of \acs{CGMF} was used with \texttt{Python} bindings, and all software used to run \acs{CGMF} and analyze results is open-source and available online under the name \texttt{ompuq} \footnote{\url{https://github.com/beykyle/omp-uq}}. An open-source code for uncertainty-quantification in few-body reaction calculations by the authors called \texttt{OSIRIS} \cite{Beyer_OSIRIS} was used as a dependency within \acs{CGMF}. \texttt{OSIRIS} accepts as input a set of global optical potential parameters, including for \ac{KDUQ} and \ac{WLH} (e.g. in a \texttt{json} file according to the format used by the \texttt{TOMFOOL} code \cite{pruitt2023uncertainty} which was used to calculate \ac{KDUQ}). Insofar as was possible, all components in the software stack created by the authors employ unit and regression testing using \texttt{pytest} \cite{pytestx.y} and \texttt{Catch2} \cite{catch2}.

The propagation of uncertainties through \acs{CGMF} was a significant computational task. Calculating fission observables for a single parameter sample (consisting of one million \acs{CGMF} histories) required roughly 30 cpu-hours on the Great Lakes cluster at the University of Michigan, using Intel(R) Xeon(R) Gold 6140 CPUs. The 300 samples required to generate a posterior predictive distribution of observables corresponding to a single potential and fissioning system therefore each required roughly 9300 cpu-hours.

\section{Results}
\label{sec:results}

\subsection{Prompt neutron observables}

The prompt neutron observables refer to those in which only the neutrons need be measured, not the fragments as well. These include neutron multiplicity and energy distributions.

\begin{figure*}[htp]
  \centering
  \quad
  \begin{subfigure}{0.48\textwidth}
  \begin{center}
    \includegraphics[width=\linewidth]{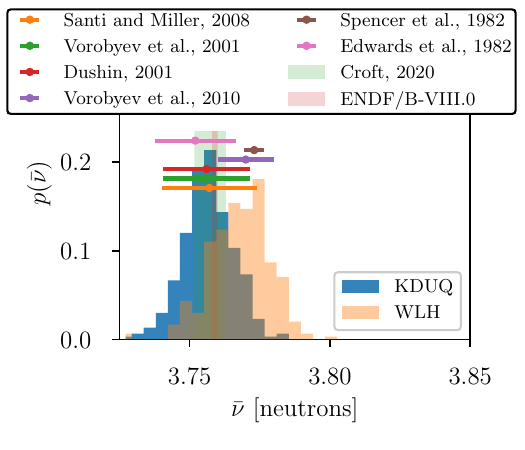}
  \end{center}
  \caption{$^{252}$Cf$(sf)$}
  \end{subfigure}
  \begin{subfigure}{0.48\textwidth}
  \begin{center}
    \includegraphics[width=\linewidth]{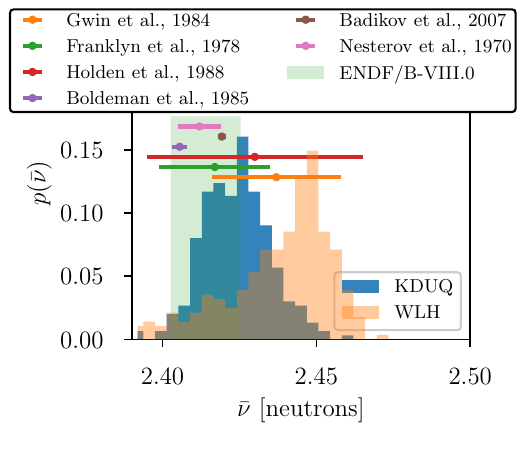}
  \end{center}
  \caption{$^{235}$U$(n_{\rm{th}},f)$}
  \end{subfigure}
  
  \caption{
  \label{fig:n}
  Distribution of average prompt neutron multiplicities, $\bar{\nu}$, produced by the uncertainty-quantified \ac{OMP}s. (a) $\bar{\nu}$  in $^{252}$Cf$(sf)$, compared to evaluations from Croft et al., \cite{CROFT2020161605}, and ENDF/B-VIII.0 \cite{brown2018endf}, as well as experimental results from \cite{doi:10.13182/NSE07-85, vorobyev2010angular, DUSHIN2004539, vorobyev2001distribution, doi:10.13182/NSE82-A18973, EDWARDS1982127}. (b) $\bar{\nu}$  in $^{235}$U$(n_{\rm{th}},f)$,
compared to the ENDF/B-VIII.0 evaluation \cite{brown2018endf}, as well as experimental results and other evaluations \cite{gwin1984measurements,franklyn1978angular,holden1988prompt,doi:10.13182/NSE85-A17133,carlson2009international,kikuchi1977neutron}. }
\end{figure*}

Figure~\ref{fig:n} displays $\bar{\nu}$, the mean prompt neutron multiplicities per fission event (inclusive of both fragments) for both \ac{OMP}s as compared to a variety of experiments and evaluations. In each of these cases, the $\sigma_{mc}$ in each ensemble was negligible compared to the parametric uncertainty of each \ac{OMP}, and are therefore not shown.  The distributions of model predictions are largely overlapping, with both models generally falling into the range of experiments and evaluations, although \ac{WLH} predicts slightly larger multiplicities, and has larger variance.

This can be interpreted from an energy budget perspective, as shown below (see e.g. Fig.~\ref{fig:pfnsacf} for example). The skewed low-multiplicity tail in Fig.~\ref{fig:n} corresponds to a skew toward softer spectra produced by \ac{WLH}. For a given initial excitation energy, less energy removed per neutron implies a larger $\bar{\nu}$. Of course, as mentioned, parameters in \acs{CGMF} relating to the fragment initial conditions are tuned to experiment assuming the default Koning-Delaroche \ac{OMP}. That is to say, we can only say that the two \ac{OMP}s are different to statistical significance, not that one is more accurate than the other.

In figures other than Fig.~\ref{fig:n}, the colored shaded regions respectively represent a credible interval of one standard deviation for each \ac{OMP} ($\sigma_{omp}$), the total shaded regions in the figures represent the total uncertainty $\sigma_t$, and the grey portions $\sigma_{mc}$ (see App.~\ref{ch:appA}). For most observables, the grey bands are negligible, indicating good convergence of the actual \ac{OMP} parameter uncertainty.

\begin{figure*}
    \centering
    \includegraphics[width=\textwidth]{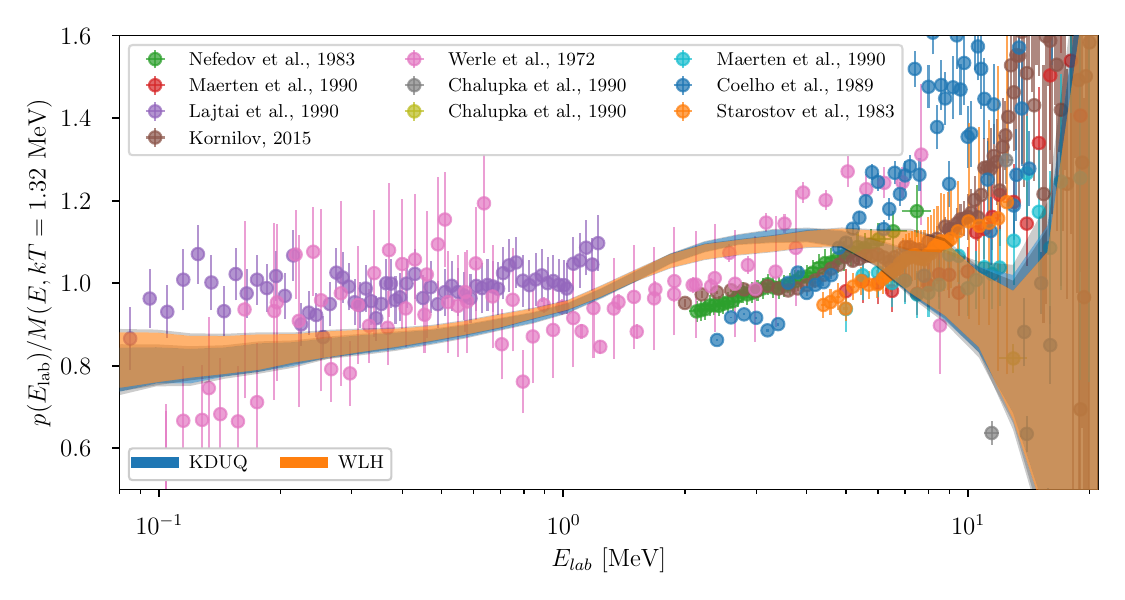}
    \caption{
      \label{fig:pfnscf}
      \ac{PFNS} in the lab frame for $^{252}$Cf$(sf)$ as a ratio to a Maxwellian at $kT = 1.32$ \unit{MeV}, compared to experimental results from \cite{marten1990252cf,nefedov1983high,lajtai1990low,kornilov2015verification,werle1972fission,chalupka1990results,coelho1989neutron,starostov1983high}.
}
\end{figure*}

\begin{figure*}
    \centering
    \includegraphics[width=\textwidth]{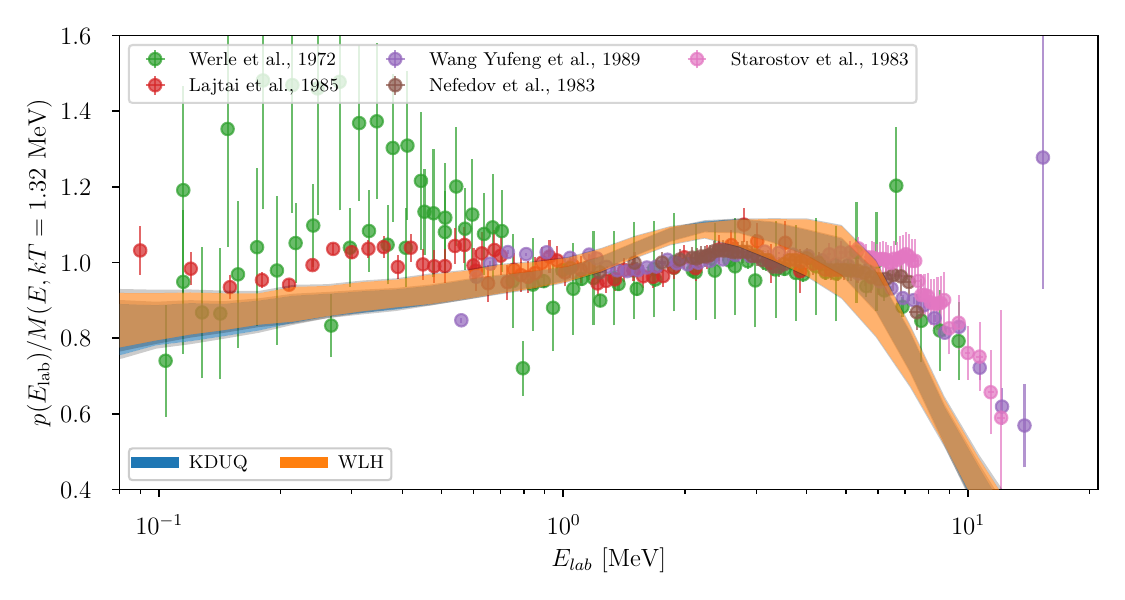}
    \caption{
      \label{fig:pfnsu}
      \ac{PFNS} in the lab frame for $^{235}$U$(n_{\rm{th}},f)$, as a ratio to a Maxwellian at $kT = 1.32$ \unit{MeV}, compared to experimental results from \cite{werle1972fission,lajtai1986energy,yufeng1992experimental,nefedov1983high,starostov1983high}.
}
\end{figure*}

Figures~\ref{fig:pfnscf} and~\ref{fig:pfnsu} display the lab-frame \ac{PFNS} for $^{252}$Cf$(sf)$ and $^{235}$U$(n_{\rm{th}},f)$ compared to a variety of measurements, all as ratios to a Maxwellian at $kT = 1.32 \unit{MeV}$. Both models roughly agree, predicting spectra that are too soft as compared to measurements, typically enhanced above 1 \unit{MeV} and de-enhanced at and above 10 \unit{MeV} relative to experiment. 
{ As mentioned, \ac{WLH} produces marginally softer spectra on average, with slightly wider distributions. Interestingly, for $^{252}$Cf$(sf)$, the uncertainty is greatly enhanced in the high outgoing energy region, but this is not the case for $^{235}$U$(n_{th},f)$, highlighting the complex mass-energy dependence of uncertainty. For $^{235}$U$(n_{th},f)$, experimental data in the high-energy region clearly demonstrate the relative softness of the spectra predicted by \acs{CGMF}, but for $^{252}$Cf$(sf)$ systematic variation between different experiments make this trend less clear}.

\subsection{Prompt neutron-fragment correlations}

\begin{figure*}
  \centering
    \begin{subfigure}{0.49\textwidth}
    \begin{center}
      \includegraphics[width=\linewidth]{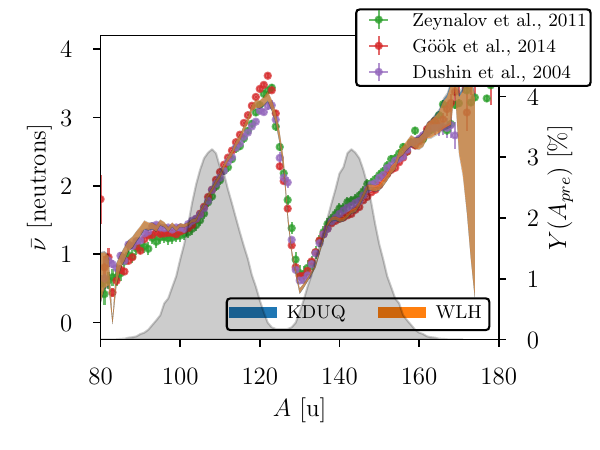}
    \end{center}
    \caption{$^{252}$Cf$(sf)$}
    \end{subfigure}
    \begin{subfigure}{0.49\textwidth}
    \begin{center}
      \includegraphics[width=\linewidth]{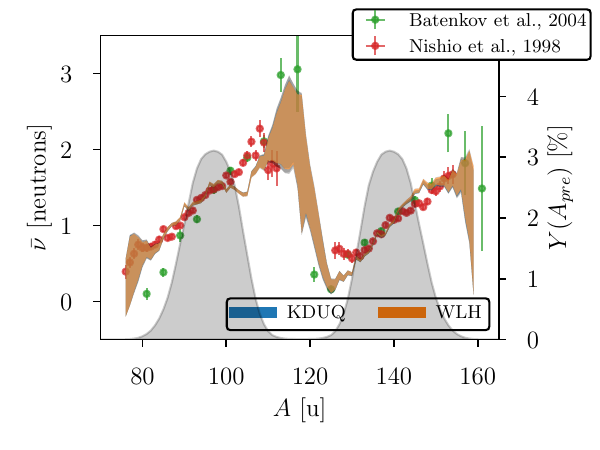}
    \end{center}
    \caption{$^{235}$U$(n_{\rm{th}},f)$}
    \end{subfigure} 
    \caption{
    \label{fig:nubarA}
  Average neutron multiplicity from fragments of a given mass number for (a) $^{252}$Cf$(sf)$ compared to experimental results from \cite{zeynalov2009neutron,BUDTZJORGENSEN1988307,bowman1963further,ding1984research,gook2014prompt,DUSHIN2004539}, with the fragment pre-emission yields from \cite{romano2010fission} overlaid on the right axis, and (b) $^{235}$U$(n_{\rm{th}},f)$ compared to experimental results from \cite{batenkov2005prompt,NISHIO1998540}, with the fragment pre-emission yields from \cite{al2020prompt} overlaid on the right axis.  
}
\end{figure*}

We turn our attention now to experiments in which neutrons and fragments were measured simultaneously. These represent a much smaller, yet highly informative, set of experiments. Neutron-fragment correlations provide detailed information about the fragments immediately after scission, i.e. the neutron multiplicity sawtooth observed by a variety of experimental groups and displayed in Fig.~\ref{fig:nubarA}. To guide the eye, the corresponding experimental pre-emission fragment mass yields $Y(A_{pre})$ are overlaid on the right axis. 

The behavior of $\bar{\nu}(A)$ indicates roughly the allocation of excitation energy ($\bar{\nu}$  being an approximate surrogate) between fragments. Both \ac{OMP}s reproduce well the expected sawtooth behavior. Multiplicities for the mass region from (roughly) $A=126$ to $A= 136$, and the corresponding light fragments from $A = 99$ to $109$ are underestimated for $^{235}$U$(n_{th},f)$, but this prediction is the same for both models and \ac{OMP} uncertainties in these mass regions are small, which suggests this effect is due to other model inputs in \acs{CGMF}. In general, this observable is not strongly correlated to the \ac{OMP} parameters, with the exceptions of the highly asymmetric (low-statistics) regions for both fissioning isotopes, and the highly symmetric region for $^{235}$U. 

\begin{figure*}
  \centering
    \begin{subfigure}{0.49\textwidth}
    \begin{center}
      \includegraphics[width=\linewidth]{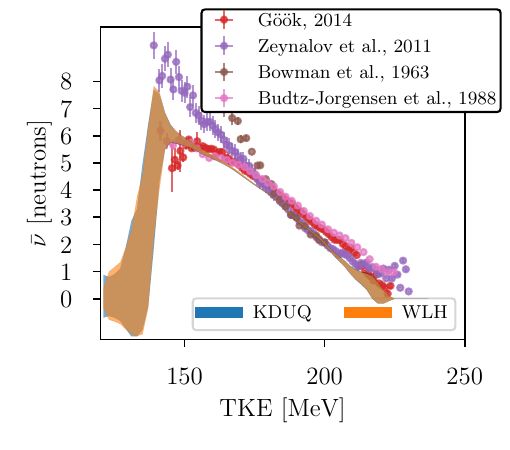}
    \end{center}
    \caption{$^{252}$Cf$(sf)$}
    \end{subfigure}
    \begin{subfigure}{0.49\textwidth}
    \begin{center}
      \includegraphics[width=\linewidth]{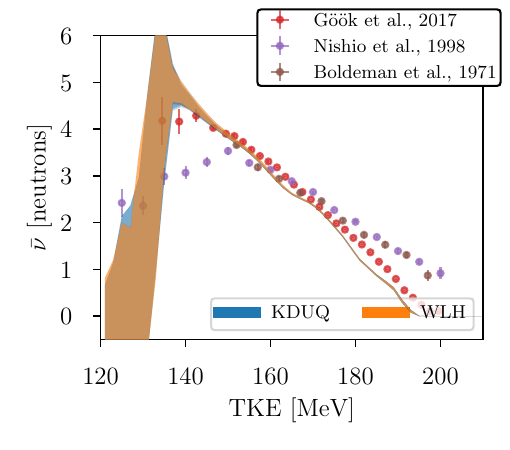}
    \end{center}
    \caption{$^{235}$U$(n_{\rm{th}},f)$}
    \end{subfigure}
    
    \caption{
    \label{fig:nubarTKE}
  Average neutron multiplicity as a function of \ac{TKE} of the fragments for (a) $^{252}$Cf$(sf)$ compared to experimental results from \cite{gook2014prompt,zeynalov2009neutron,bowman1963further,BUDTZJORGENSEN1988307}, and (b) $^{235}$U$(n_{\rm{th}},f)$ compared to experimental results from \cite{gook2018prompt,NISHIO1998540,boldeman1971prompt}}
\end{figure*}

Figure~\ref{fig:nubarTKE} displays the mean single-fragment $\bar{\nu}$ as a function of the \ac{TKE} of both fragments, compared to a variety of measurements. The approximately linear decrease in neutron multiplicity as a function of \ac{TKE} indicates the strong correlation between excitation energy and multiplicity, and is an important signature in understanding the energy cost of evaporating neutrons from fully-accelerated fragments. For both fissioning isotopes, this observables does not exhibit much sensitivity to the to the parameters in either model, except for in the low \ac{TKE}, highly symmetric, fission region. This region corresponds to highly deformed fragments post-scission, and, therefore, excitation energy dominated systems during de-excitation. Both models reproduce experiment well, especially the most recent measurement by G\"o\"ok \cite{gook2014prompt}. The other experiments disagree for \ac{TKE} $< 160$ \unit{MeV}. 

The lack of correlation to the \ac{OMP} of $\langle \nu | A\rangle$ and $\langle \nu | \ac{TKE} \rangle$ indicates the utility of these observables for inclusion as constraints of the global optimization of model inputs unrelated to the \ac{OMP}, especially those related to scission, e.g. the excitation energy partitioning. In general, the observables relating to the correlations between fragment mass/\ac{TKE} and neutron multiplicity do not show strong sensitivity to the \ac{OMP} parameters, and both models agree, with much smaller model uncertainties and differences between the two models than differences between experiments and between model and experiment.

\begin{figure*}
    
  \centering
  \begin{subfigure}{0.49\textwidth}
  \begin{center}
    \includegraphics[width=\linewidth]{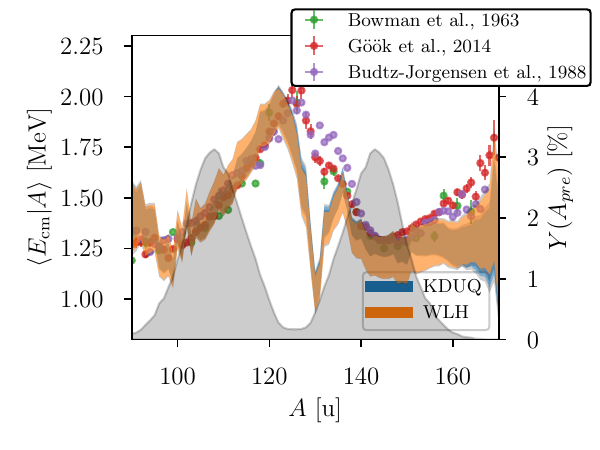}
  \end{center}
  \caption{$^{252}$Cf$(sf)$}
  \end{subfigure}
  \begin{subfigure}{0.49\textwidth}
  \begin{center}
    \includegraphics[width=\linewidth]{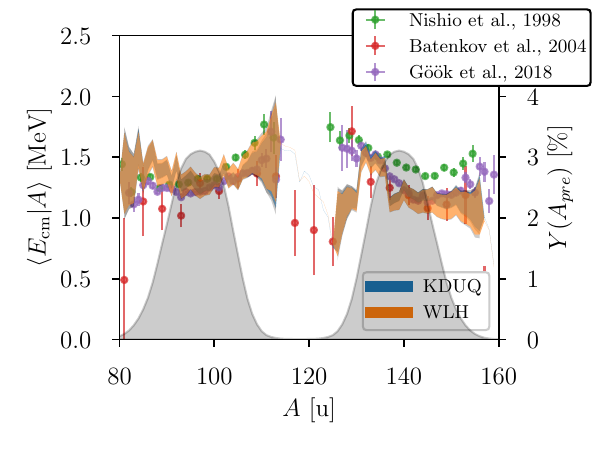}
  \end{center}
  \caption{$^{235}$U$(n_{\rm{th}},f)$}
  \end{subfigure}

    \caption{
      Average neutron energy in the \ac{COM} frame emitted from fragments of a given mass number for (a) $^{252}$Cf$(sf)$, compared to experimental results from \cite{bowman1963further,gook2014prompt, BUDTZJORGENSEN1988307}, with the fragment pre-emission yields from \cite{romano2010fission} overlaid on the right axis, and (b) $^{235}$U$(n_{\rm{th}},f)$ \cite{NISHIO1998540,batenkov2005prompt,gook2018prompt}, with the fragment pre-emission yields from \cite{al2020prompt} overlaid on the right axis. 
 }
    \label{fig:encomA}
\end{figure*}

\begin{figure*}
  \begin{subfigure}{0.49\textwidth}
  \begin{center}
    \includegraphics[width=\linewidth]{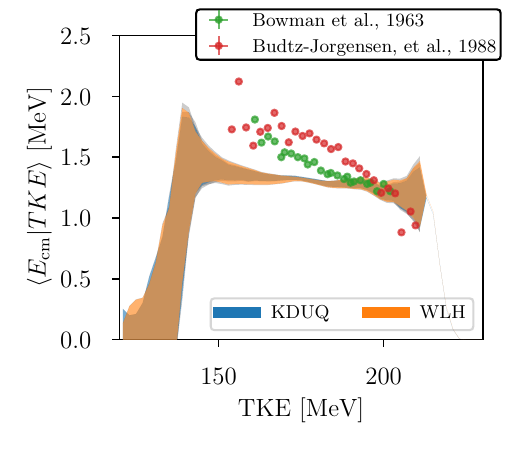}
  \end{center}
  \caption{$^{252}$Cf$(sf)$}
  \end{subfigure}
  \begin{subfigure}{0.49\textwidth}
  \begin{center}
    \includegraphics[width=\linewidth]{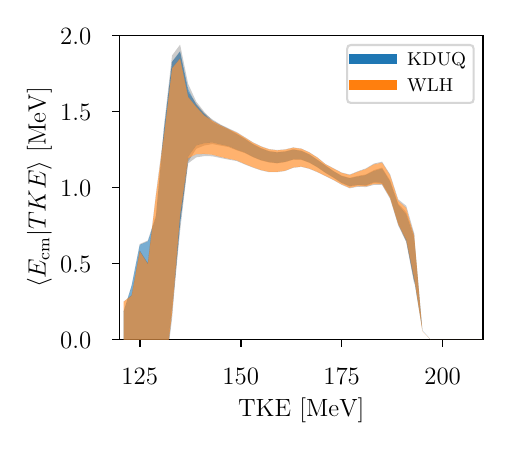}
  \end{center}
  \caption{$^{235}$U$(n_{\rm{th}},f)$}
  \end{subfigure}
    
  \caption{Average neutron energy in the \ac{COM} frame, as a function of the \ac{TKE} of the fragment pair, for (a) $^{252}$Cf$(sf)$ compared to experimental results from \cite{bowman1963further,BUDTZJORGENSEN1988307}, and (b). $^{235}$U$(n_{\rm{th}},f)$ }
    \label{fig:encomTKE}
\end{figure*}

We now turn our attention to observables relating to the correlation between fragment mass/\ac{TKE} and mean neutron energy. These experimental comparisons typically required kinematic correction for \ac{COM}-frame \ac{PFNS}. In particular, as described in \cite{gook2014prompt}, only neutrons with lab-frame energy larger than the pre-emission kinetic energy per nucleon of the emitting fragment were measured. For comparison to this experiment, these cuts were applied event by event to the generated histories. A similar cut to compare to data from \cite{bowman1963further} was determined to have a negligible effect when applied to \acs{CGMF} histories, and was subsequently ignored. Additionally, mass reconstruction in the relevant experiments has an uncertainty with a standard deviation of roughly 2-4 mass units.

Figure~\ref{fig:encomA} displays the mean neutron energy as a function of fragment mass. 
{Both models predict spectra that are too soft, however this is not the case uniformly across mass number; the mass dependent behavior of each \ac{OMP} is slightly different, with \ac{WLH} predicting slightly softer spectra than \ac{KDUQ} in the heavy mass region and slightly harder in the light mass region. This effect is especially pronounced in the heavy fragment region for $^{252}$Cf, where the sensitivity of the mean neutron energy to the \ac{OMP} is pronounced, and the energy is systematically under predicted.  For $^{235}$U$(n_{th},f)$, $\sigma_{mc}$ was larger for this observable in the low-statistics $116 < A < 126$ mass region, so no \ac{OMP} confidence interval could be extracted in this region.  }  

Both models exhibit poor agreement, for both fissioning isotopes, (roughly) in the mass region $126 < A < 136$, which corresponds to the heavy fragment in the pair being at or near the $^{132}$Sn double shell closure. Due to the presence of the shell closure, both fissioning systems are similarly likely to populate fragments into this mass range, with $^{235}$U having slightly larger yields here. In the case of $^{252}$Cf$(sf)$, this is near-symmetric fission, while, for $^{235}$U$(n_{th},f)$ this corresponds to slightly asymmetric fission. Interestingly, in $^{235}$U$(n_{th},f)$, this is the same region in which multiplicities are underestimated in Fig.~\ref{fig:nubarA}.  Taken together, these anomalies potentially indicate deficiency in the model for excitation energy sharing for fragments near the shell closure.

Figure~\ref{fig:encomTKE} displays the mean neutron energy emitted from individual fragments as a function of the \ac{TKE} of both fragments. Both models agree to within uncertainty. It is worth noting the disagreement obtained with experiment, further investigation into the difference between \acs{CGMF} predictions and experimental measurements of $\langle E | \ac{TKE} \rangle$ should clarify this issue.

\subsection{Differential prompt neutron-fragment correlations}

Finally, we consider these same observable differentiated along neutron energy and \ac{TKE}. For each observable, we look at a set of selected mass pairs, showing the light fragment in the pair in the left columns of Figs.~\ref{fig:pfnsacf},\ref{fig:pfnsau},\ref{fig:nubaratkecf},\ref{fig:nubaratkeu},and \ref{fig:encomatkecf}, and the heavy fragment in the right columns. As the mass resolution typical of these experiments is greater than 1 mass unit, we indicate the mass centroid of the data in each panel as approximate, e.g. by $A \sim 100$.

The mass dependence of the \ac{PFNS} is explored further in Figs.~\ref{fig:pfnsacf} and ~\ref{fig:pfnsau}, where single-fragment \ac{COM}-frame \ac{PFNS}, $p(E_{cm} | A)$ , are compared to \cite{gook2014prompt}, for a few selected mass number pairs.  
{Generally, both \ac{OMP}s agree, with the uncertainties in both models being on the order of the experimental uncertainties. }

Interestingly, the spectra indicate a strong deviation from Maxwellian behavior for some fragments with some experiments having more than 5 times the number of neutrons near $\sim 10$ \unit{MeV} than either model predicts. 
{
  Although the experimental uncertainties are large, they are generally well outside the model confidence intervals, indicating this disagreement cannot be explained by the optical potential. Indeed, it seems unlikely that a Hauser-Feshbach model describing neutron evaporation from a fully-accelerated fragment would predict many neutrons near 10 \unit{MeV} in the \ac{COM} frame. Further, this effect persists to different degrees across all masses considered, which suggests it is unlikely to  be explained by some missing mass-dependent physics in fragment formation (e.g. excitation energy sharing), as was suggested for the anomalies in the $126 < A < 136$ mass region in the previous section. Indeed, even the near-shell-closure fragments have these high energy tails, indicating that this effect likely requires further explanation. Additionally, multiple different experiments in $^{235}$U$(n_{th},f)$ corroborate this effect, suggesting it is not an issue due to systematics. Finally, the fact that this feature is present in the thermal-induced and spontaneously fissioning systems considered here indicates that these neutrons are not emitted before scission, i.e. as pre-equilibrium neutrons or $N^{th}$-chance fission. So where are these neutrons coming from?
}

{
  One potential explanation could be non-statistical effects like very strong coupling of continuum states in the fragment to low-lying discrete states in the residual core, which are not included in \acs{CGMF} due to the lack of knowledge of the structure of these exotic nuclei. This possibility is explored in \cite{PhysRevC.104.014611}, but no such analysis has been done which compares to these experiments. Another possibility is the presence of scission neutrons, which wasn't taken into
account in the \ac{COM}-frame reconstruction in either experiment (and indeed, would be very difficult to do). The presence of scission neutrons has been recently suggested by time-dependent simulations of scission dynamics \cite{PhysRevLett.132.242501}.} 

{ 
  Either way, this experimental signature does not seem to have any explanation within a Hauser-Feshbach model, and further precision measurement of event-by-event fission fragments in correlation with neutrons could shed further light. In particular, fission arm spectrometers, like the SPIDER detector \cite{meierbachtol2015spider}, in coincidence with a scintillator
array, could provide event-by-event correlations between neutron energy and fragment $A$, $Z$, and \ac{TKE}, as well the angle between the fission axis and the emitted neutrons, which would shed light on the origin of the high-energy tails.
}

Returning to effects that \textit{can} be explained by the \ac{OMP}, these fragment-correlated neutron spectra exhibit larger sensitivities to \ac{OMP} parameters than other explored in this work, potentially indicating their utility in constraining an \ac{OMP}. Conversely, this same sensitivity makes them unsuitable for calibrating other, non-\ac{OMP}, model inputs in \acs{CGMF}.

Figures~\ref{fig:nubaratkecf} and \ref{fig:nubaratkeu} display the mean single-fragment multiplicity conditional upon mass number and \ac{TKE} of the fragment pair, $\langle \nu|TKE,A \rangle$, for a few selected mass number pairs. We see high degree of sensitivity of $\langle \nu|TKE,A \rangle$  to the \ac{OMP} in the low \ac{TKE} symmetric mass region. The experimental data is reproduced well for the most part, with the exception of systematic over/under-estimation for fragment pairs; e.g.
$A\sim 104, 131$ in $^{235}$U. Both models are essentially identical for these observables. Because of this low sensitivity to the \ac{OMP}, and the physical meaning of $d\ac{TKE}/d\nu$ (discussed in the previous section and in \cite{gook2014prompt}), make this set of observables ideal for constraining the excitation energy sharing in scission. 

Figure~\ref{fig:encomatkecf} displays the mean neutron energy conditional upon fragment mass and \ac{TKE}, $\langle E_{cm} | A, TKE \rangle$. While this observable could in principle be extracted from most of the other similar experiments; e.g.~\cite{gook2014prompt,gook2018prompt,BUDTZJORGENSEN1988307}, it is only reported by Bowman et al. \cite{bowman1963further} and only for $^{252}$Cf. As this is the most sensitive observable to the optical model, we recommend further experimental study in
this direction. 

{The overall shape as a function of \ac{TKE} was roughly the same for both models.} Although the experimental data points are sparse, the agreement is reasonable with the exception of $A = 131$, in which it is off by up to an \unit{MeV} in the region of \ac{TKE} $\sim 200$ \unit{MeV}. This is well outside of the parametric model uncertainty. The proximity to the shell-closure suggests this is related to the anomalies in the same mass region discussed in the
previous section. The limited experimental data points in Fig.~\ref{fig:encomatkecf} suggest that the neutron energy removed does not follow as strict of a correlation with excitation energy as neutron multiplicity, and there may be some structure worth exploring. 

\begin{figure}
\begin{center}
  \includegraphics[width=0.5\textwidth]{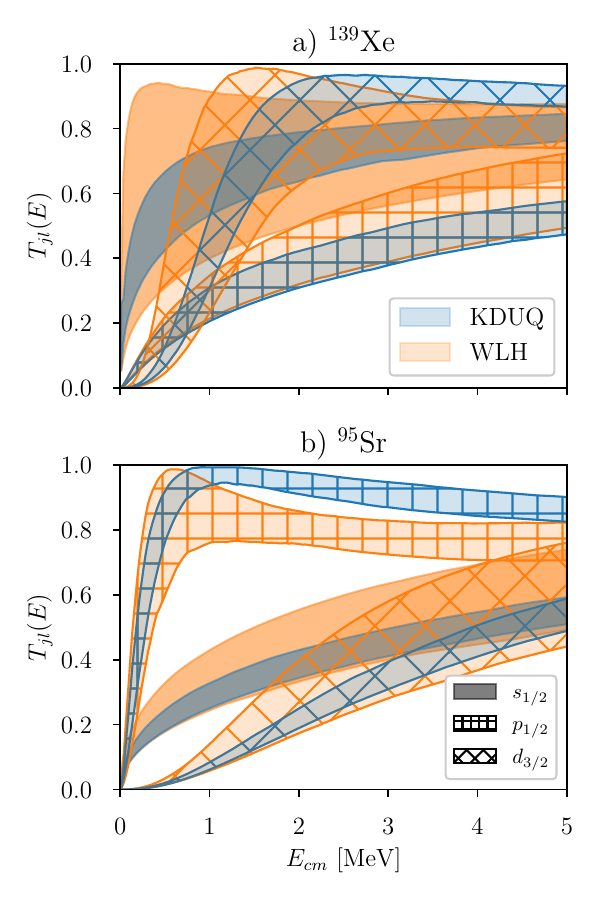}
\end{center}
\caption{68\% credible intervals for transmission coefficients for three partial waves in two representative fission fragments predicted by \ac{KDUQ} and \ac{WLH} \label{fig:tcoeff}}
\end{figure}

\begin{figure*}[hbtp]
\begin{center}
  \includegraphics[width=\textwidth]{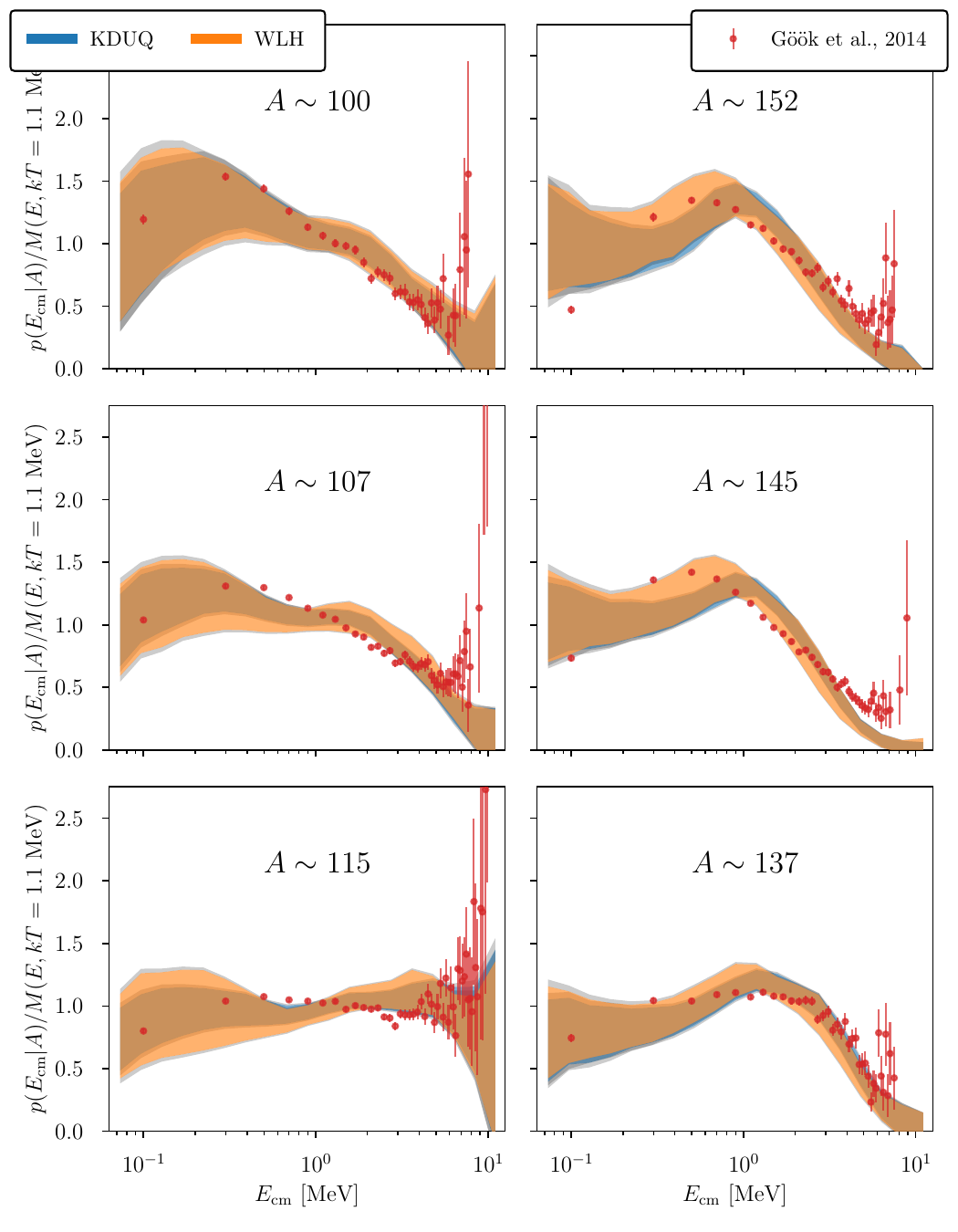}
\end{center}
\caption{\label{fig:pfnsacf} \ac{PFNS} ratio to a Maxwellian with $kT = 1.1$ \unit{MeV}, conditional on fragment mass, for $^{252}$Cf$(sf)$ compared to experimental data from \cite{gook2014prompt}.}
\end{figure*}

\begin{figure*}[hbtp]
\begin{center}
  \includegraphics[width=\textwidth]{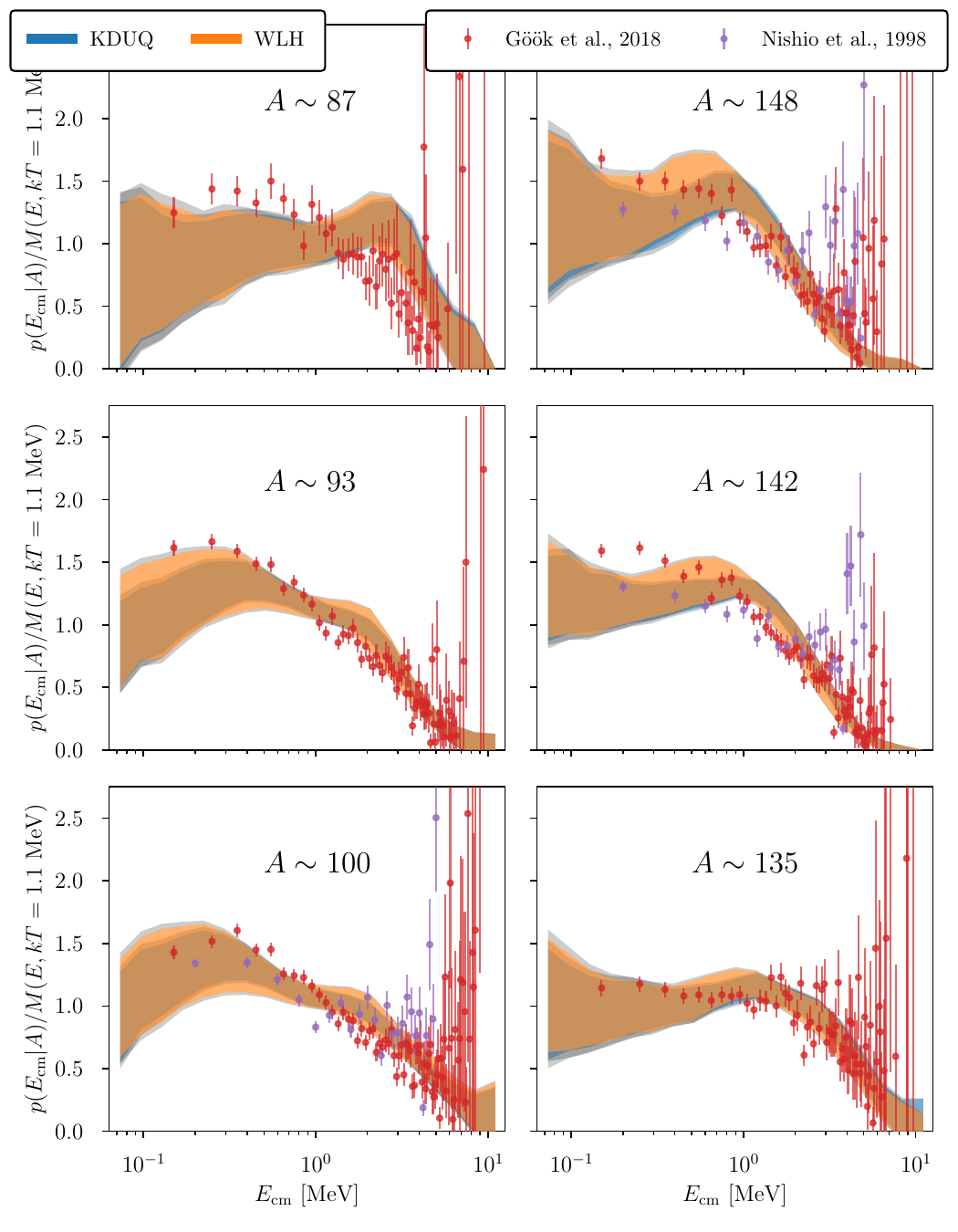}
\end{center}
\caption{\label{fig:pfnsau}
Same as Fig.~\ref{fig:pfnscf} for $^{235}$U$(n_{th},f)$ compared to experimental data from \cite{gook2018prompt,NISHIO1998540}.}
\end{figure*}

\begin{figure*}
\begin{center}
  \includegraphics[width=\textwidth]{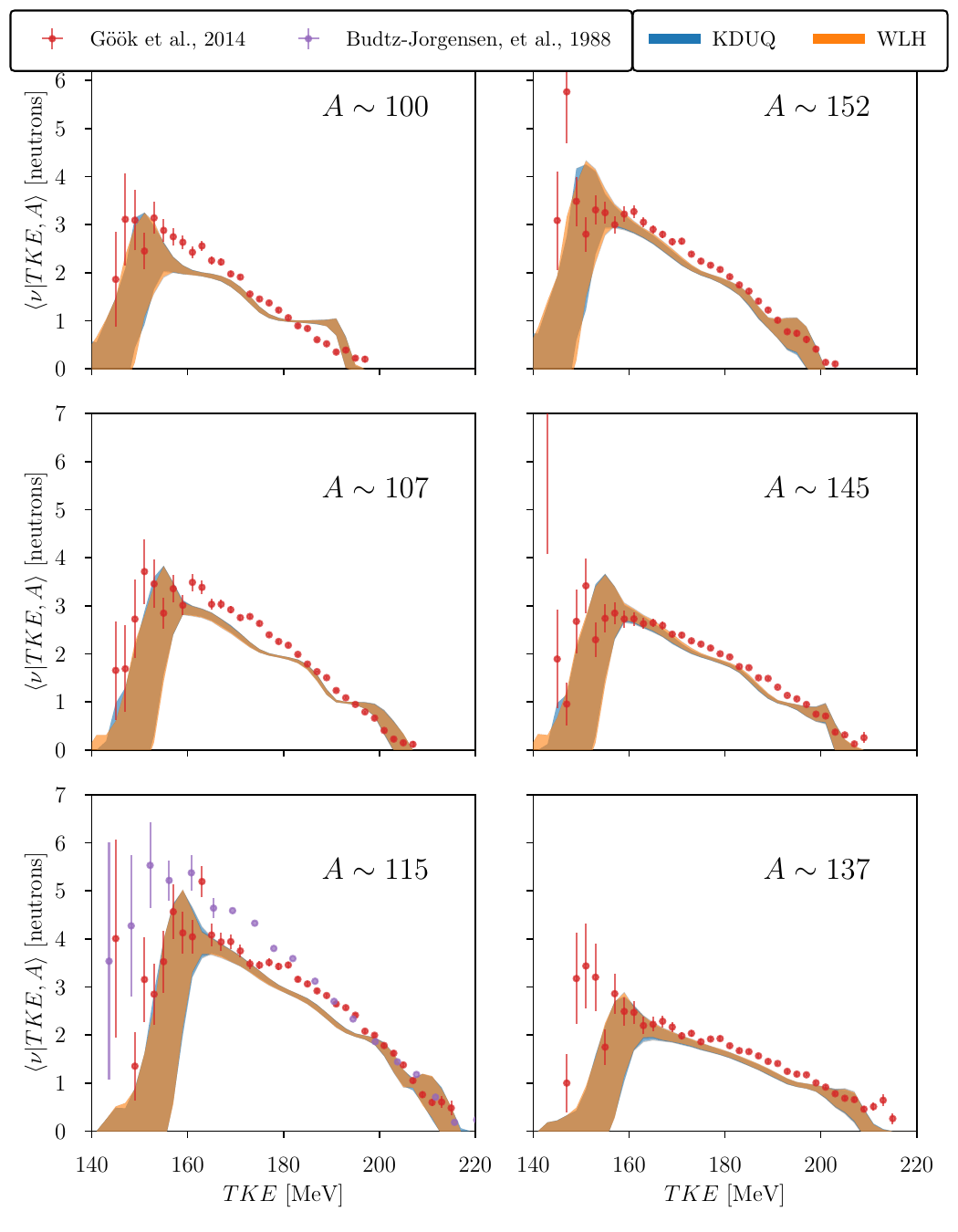}
\end{center}
\caption{\label{fig:nubaratkecf} Single-fragment neutron multiplicity as a function of the \ac{TKE} of both fragments, and conditional on fragment mass, for $^{252}$Cf$(sf)$ compared to experimental data from \cite{gook2014prompt,BUDTZJORGENSEN1988307}.}
\end{figure*}

\begin{figure*}
\begin{center}
  \includegraphics[width=\textwidth]{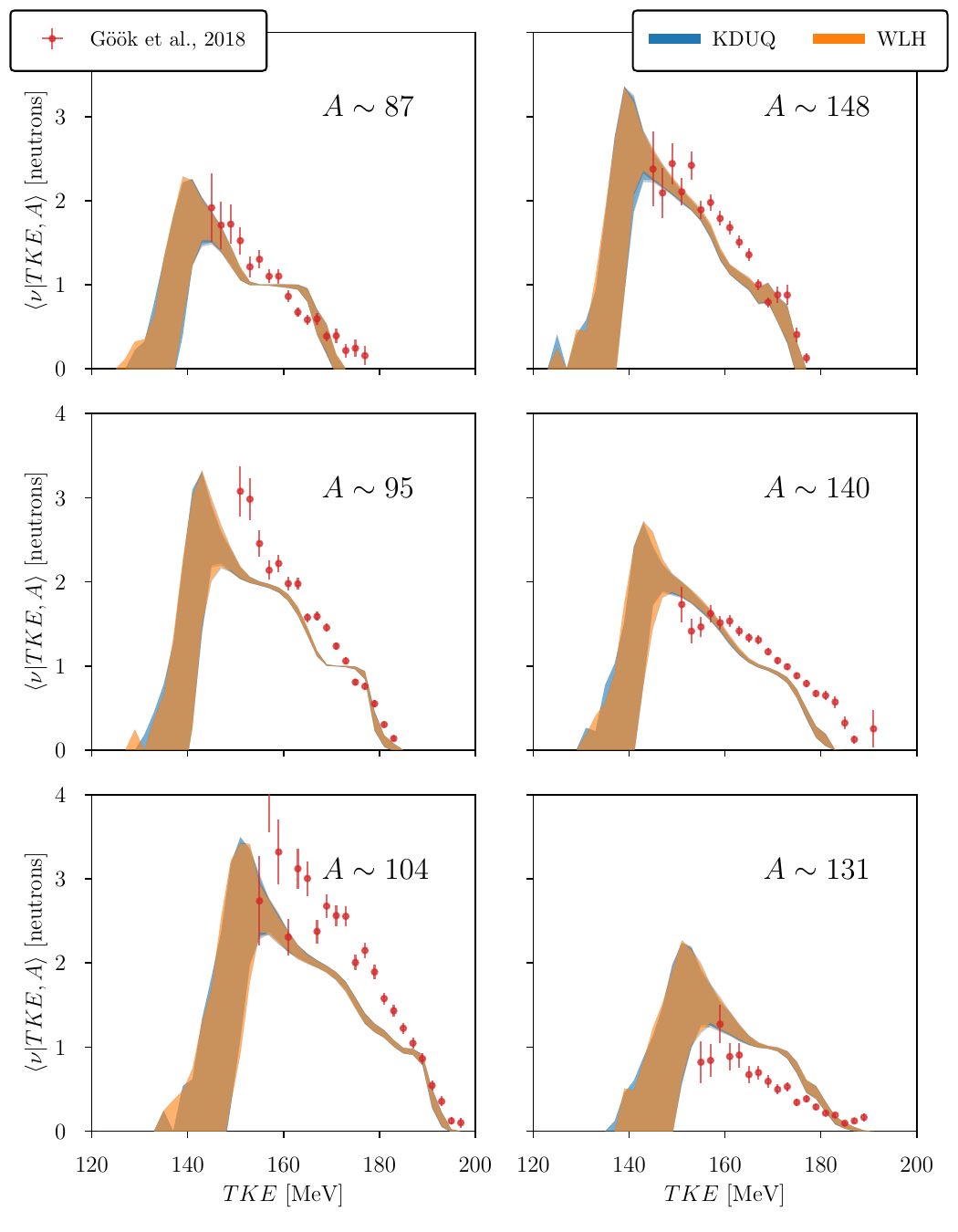}
\end{center}
\caption{\label{fig:nubaratkeu} Same as Fig.~\ref{fig:nubaratkecf} for $^{235}$U$(n_{th},f)$ compared to experimental data from \cite{gook2014prompt}.}
\end{figure*}

\begin{figure*}
\begin{center}
  \includegraphics[width=\textwidth]{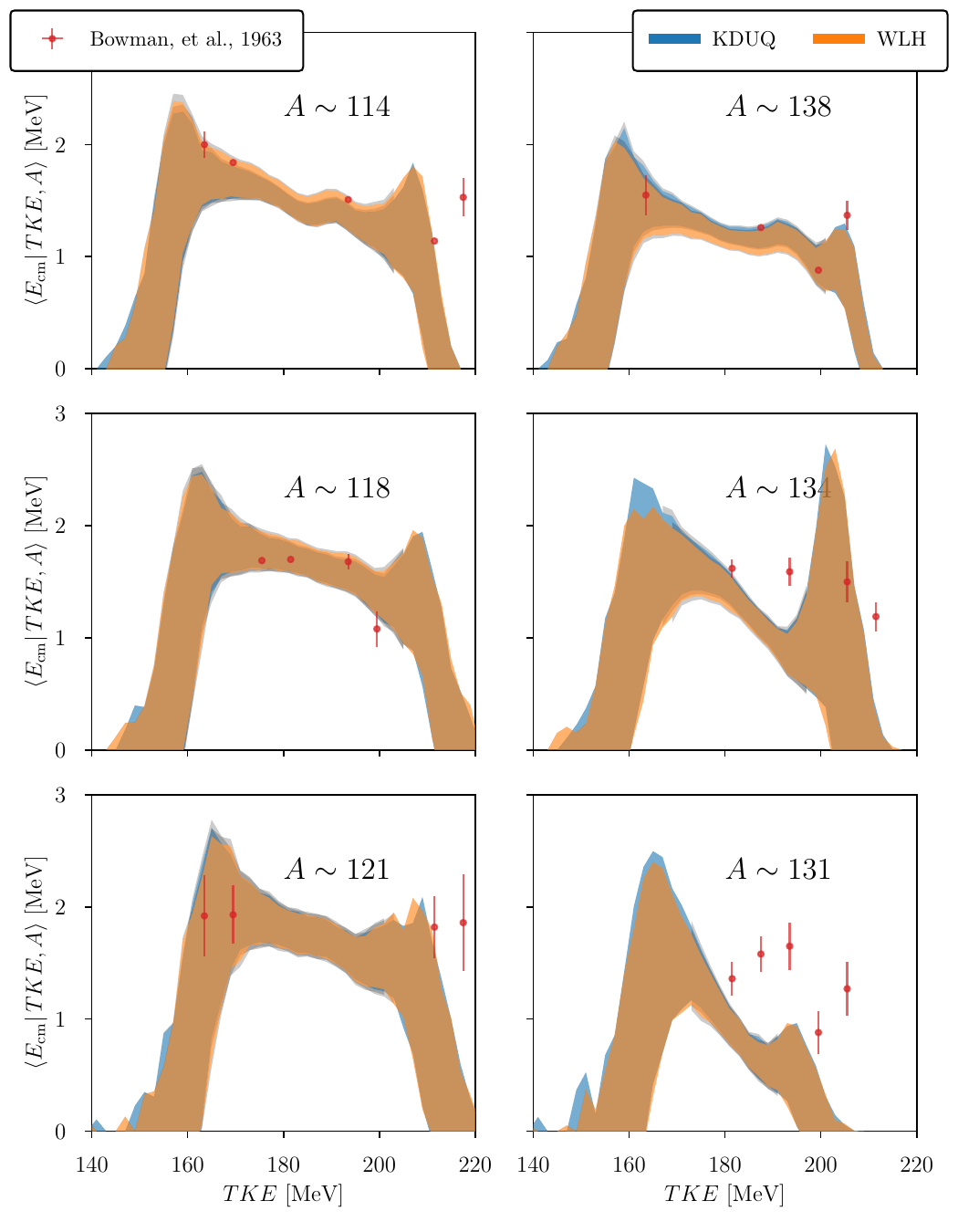}
\end{center}
\caption{\label{fig:encomatkecf} Average \ac{COM}-frame neutron energy as a function of the \ac{TKE} of both fragments, and conditional on fragment mass, for $^{252}$Cf$(sf)$ compared to experimental data from \cite{bowman1963further}.}
\end{figure*}

\subsection{Transmission coefficients}

The only effect the \ac{OMP} can have on fission fragments is by altering the probability of de-excitation pathways via the transmission coefficients in Eq.~\eqref{eq:HFfull}. Fig.~\ref{fig:tcoeff} displays confidence intervals for transmission coefficients predicted by the two models, for a representative light and heavy fragment. Clearly, within the fission energy range of interest, both models agree to within uncertainties for the most part. Where they don't (e.g. $p_{1/2}$-wave in $^{95}$Sr above 1 \unit{MeV}), they have roughly the same shape. Transmission coefficients are renormalized in Eq.~\eqref{eq:HFfull}, so this similarity in shape between models, and the enhanced uncertainties in \ac{WLH} along with its slight shift to lower energies, schematically explains how the model predictions in this work compare.

The representative fragments in Fig. ~\ref{fig:tcoeff} were chosen to recreate Fig 2. in the supplemental material of \cite{stetcu2021angular}. The argument in this work is that the optical potential in a Hauser-Feshbach calculation predicts enhanced angular momentum removal by neutrons ($> 1.5 \hbar$), contradicting the typical assumption, which is schematically explained by the enhancement of higher partial wave transmission coefficients even at lower energy. This behavior is reproduced by both models, supporting the claims made in \cite{stetcu2021angular}.

\section{Conclusions and future work}
\label{sec:conclusion}

We have shown that, while neutron-fragment correlated fission observables are sensitive to \acf{OMP} uncertainties, both model forms considered here produced roughly the same predictions for the observables we considered. In particular, neutron energy spectra are more sensitive than neutron multiplicities, the latter of which are only slightly sensitive due to energy budget considerations, which is a second-order effect and not relevant given the size of experimental uncertainties. On the other hand, neutron energy spectra, when differentiated on mass and \ac{TKE}, show significant sensitivities to the parameters of individual \ac{OMP}s. The similarity in the energy dependence of transmission coefficients produced by the two models serves to explain how they compare in fission observables. 

Based on the results reported here, we discuss two avenues of future work: improving the description of the state of the fragments following scission in the context of the experiment-model disagreements highlighted in this work, and improving the description of the optical potential used for de-excitation of the fission fragments.

\subsection{Improving the description of scission}
 
Of particular interest is the multiplicity and energy spectra of neutrons emitted in the mass region near the $^{132}$Sn shell closure. The strong disagreement between experiment and theory cannot be explained purely by the uncertainty in the \ac{OMP} priors. Further study of fission fragment initial conditions, informed by experiment, or microscopic time-dependent mean-field calculations of scission (e.g. \cite{PhysRevLett.132.242501}), may shed light on the role of excitation energy sharing in this region. 

Future work should use experimental $\langle \nu | A, \ac{TKE} \rangle$ to calibrate fragment temperature ratios governing excitation energy sharing in \acs{CGMF}, as these observables are minimally sensitive to the optical potential, and are not reproduced by \acs{CGMF} for fragment masses near the $^{132}$Sn shell closure.  In particular, measurements of mass-dependent neutron energy spectra provide a (admittedly model-dependent) measurement of fragment temperatures \cite{BUDTZJORGENSEN1988307}. Using these as Bayesian priors in such a calibration, or incorporating them as a constraint in the likelihood function, should be investigated. The ability to leverage surrogate models to do rapid uncertainty quantification of fission observables  (e.g. \cite{daningburg2024novel}) could make this much more feasible. 

{Finally, anomalously high energy neutrons are observed in multiple fragment-correlated experiments, across both fissioning isotopes studied, and over all fragment masses. These neutrons, some nearly 10 \unit{MeV} in the neutron-fragment \ac{COM}-frame, are well outside of the \ac{OMP} uncertainties calculated here. The source of these neutrons is unknown. We postulate that they may be scission neutrons, or due to non-statistical effects in the de-excitation not accounted for by  \acs{CGMF}. Future work should include microscopic distributions of scission neutrons
predicted by \cite{PhysRevLett.132.242501} in \acs{CGMF} for comparison to these experiments.}

In particular, precision fragment-neutron correlated experiments using time-of-flight fission arm spectrometers promise the most precise fission-fragment measurements to date \cite{meierbachtol2015spider}. Performing such a measurement with event-by-event correlation with prompt neutrons would provide a new and rich set of observables to validate fission models. These measurements would shed light on the shell-closure and high-energy neutron anomalies discussed, and the state of fragments immediately post-scission.

\subsection{Improving the optical potential}

We have shown that observables like $\langle E_{cm} | A , \ac{TKE} \rangle$ may be promising for directly calibrating an \ac{OMP}, once fragment initial conditions are taken care of. Especially in the case of heavy fission fragments, we expect low-lying collective excitations to arise in the form of rotational modes. Explicitly including these couplings in the form of a coupled-channels \ac{OMP} would, in principle, provide a better description, but this has not been done in the case of fission fragment de-excitation due to computational constraints and lack of information about excited states in neutron-rich nuclei. 
{One possibility is the extension of Koning-Delaroche to deformed nuclei \cite{nobre2015derivation}. Extending this work to include uncertainty-quantification and propagating it into fission observables through a Hauser-Feshbach code represents a significant, yet worthwhile goal. }As this may not be computationally tractable within \ac{MCHF}, emulators capable of significant speed up to the calculation of transmission coefficients are being explored and developed (e.g. \cite{ROSE2023,beyer2024ex}). 

{
Aside from the isospin dependence, both models utilized in this work had ambiguities in their energy dependence. Both depend on the lab-frame energy, while the \ac{COM}-frame energy is used for \ac{CN} de-excitation in Hauser-Feshbach. Additionally, \ac{KDUQ} also depends on the Fermi energy of the residual core, which it parameterizes as having an $A$, but no $Z$ dependence. Near the valley of stability this is reasonable, but the neutron Fermi energy should be decreased for neutron-rich fission fragments, which would shift the energy dependence relative to what is predicted here. 
}

{
  Future work could explore the predictions for the energy dependence of transmission coefficients in the fission fragment region by other optical potentials, which may have different asymmetry and energy dependence. If behavior differing significantly from the models considered here is predicted, it would be worth repeating this work to understand the impact on the fission observables highlighted. 
}

{
A promising model is the Bruy\`{e}res Jeukenne-Lejeune-Mahaux (JLMB) semi-microscopic folding model, which is Lane-consistent \cite{PhysRevC.58.1118,PhysRevC.58.1118}. This has been extended to statically deformed nuclei \cite{bauge1999jlm}, using mean-field nucleon densities, and applied to a wide variety of reactions \cite{hilaire2016nuclear,dupuis2015progress}. A particular advantage of these interactions, based on Gogny forces \cite{decharge1980hartree}, is that these forces can be consistently applied as ingredients in modeling all parts of the fission process, from the entrance channel, to potential energy surfaces, and finally to the fragment de-excitation. Challenging this model in such a way, with uncertainty quantification, presents another significant, yet worthwhile effort. 
}

{
  Even still, these models have the same limitation as \ac{WLH}, namely, the reliance on the Bruckner-Hartree-Fock (BHF) folding model. This approximation is limited in its description of collective correlations, which leads to a lack of imaginary strength at the surface. Future work going beyond the BHF approximation using a Particle-Vibration Coupling (PVC) could yield a useful microscopic description of the optical potential \cite{ring2009particle}. Alternatively, potentials which
  use the Skyrme interaction and \acf{MBPT} to construct a nuclear matter potential (see e.g. \cite{shen2009isospin}) are able to produce imaginary surface strength due to the current-density and gradient terms in the Skyrme interaction, and are analytic, making it easier to quantity their uncertainties.
}

\section{Acknowledgements}

The authors would like to thank Stefano Marin for useful discussions about fission fragments and neutron evaporation, and Aaron Tumulak for discussions about Monte Carlo estimators. 

This work was funded by the Consortium for Monitoring, Technology, and Verification under Department of Energy National Nuclear Security Administration award number DE-NA0003920, performed under the auspices of the U.S. Department of Energy by Los Alamos National Laboratory under Contract 89233218CNA000001 and under the auspices of the U.S. Department of Energy by Lawrence Livermore National Laboratory under Contract DE-AC52-07NA27344. This work was also supported by the Laboratory Directed Research and Development program of Los Alamos National Laboratory under project number 20220532ECR.

\appendix
\section{Uncertainty propagation with Monte Carlo}
\label{ch:appA}

The spread of the mean in each observable over all parameter samples reflects the sum of two uncertainties: the intrinsic parametric uncertainty of each \ac{OMP}, and the inherent uncertainty due to the limited number of Monte Carlo histories per parameter sample. These uncertainties are assumed to be uncorrelated, so that the total uncertainty $\sigma_t$ across the ensemble of \ac{OMP} samples was taken as the quadrature sum of the \ac{MCHF} uncertainty and the parametric uncertainty of the \ac{OMP}. The latter was estimated as

\begin{equation}
  \label{eq:unc}
  \sigma_{omp} = \sqrt{\sigma_t^2 - \sigma_{mc}^2 }.
\end{equation}
 
For a given ensemble of histories corresponding to \ac{OMP} parameter sample $i$, a Monte Carlo uncertainty $\sigma^i_{mc}$ was calculated for each observable.  For observables that correspond to the mean of a distribution over events (e.g. average prompt neutron multiplicity $\bar{\nu}$), $\sigma^i_{mc}$ corresponded to the \ac{SEM}, and was estimated directly from the second moment of the distribution of the observable over all histories in the ensemble:

\begin{equation}
  \sigma^i_{mc} = \sqrt{\frac{1}{N} \sum_n (k^i_n)^2 - \frac{1}{N^2} \left( \sum_n k^i_n \right)^2 },
\end{equation}

\noindent
where $k^i_n$ represents the score for the observable of interest in the $n^{\rm{th}}$ (out of $N$) histories of the $i^{\rm{th}}$ sample. 

Observables corresponding to distributions (e.g. \ac{PFNS}), were constructed by histogramming over ensemble histories, in which case the mean of an observable in a bin corresponded to the probability density of $k^i_n$ falling into that bin in a given history. For $m$ times out of $N$ histories in a bin in phase space with width $\Delta x$, the mean is $m / (N \Delta x)$. In this case, $\sigma^i_{mc}$ was estimated in each histogram bin according to a Binomial distribution:

\begin{equation}
  \sigma^i_{mc} = \frac{1}{N \Delta X}\sqrt{ 1 - \frac{m}{N} }.
\end{equation}

\noindent
In either case, $\sigma_{mc}$ for the observable was then estimated by averaging $\sigma^i_{mc}$ over all ensembles. 

\begin{acronym}
    \acro{OMP}{optical model potential}
    \acro{CN}{compound nucleus}
    \acro{CGMF}{\href{https://github.com/beykyle/cgmf}{\texttt{Cascade Gamma Multiplicity Fission}}}
    \acro{COM}{center-of-mass}
    \acro{WFC}{width fluctuation correction}
    \acro{MCHF}{Monte Carlo Hauser-Feshbach}
    \acro{HF}{Hauser-Feshbach}
    \acro{TKE}{total kinetic energy}
    \acro{TXE}{total excitation energy}
    \acro{PFNS}{prompt fission neutron spectrum}
    \acro{WLH}{Whitehead-Lim-Holt}
    \acro{KD}{Koning-Delaroche}
    \acro{KDUQ}{Koning-Delaroche uncertainty quantified}
    \acro{KCK}{Kawano-Chiba-Koura}
    \acro{DWBA}{distorted-wave Born approximation}
    \acro{FPY}{fission product yield}
    \acro{SEM}{standard error in the mean}
    \acro{MBPT}{many-body perturbation theory}
    \acro{QRPA}{quasi-particle random phase approximation}
    \acro{xeft}[$\chi$EFT]{Chiral effective field theory}
    \acro{QCD}{quantum chromodynamics}
    \acro{I-LDA}{improved local density approximation}
\end{acronym}

\bibliographystyle{apsrev4-1}
\bibliography{ffuq}

\end{document}